\documentclass[aps,twocolumn]{revtex4}

\usepackage{graphics}
\usepackage{epsfig}
\usepackage{bm}
\usepackage{amsmath,amssymb,framed,color}
\usepackage{wasysym}

\def\vb#1{\mbox{\boldmath$#1$}}
\def\pd#1#2{\frac{\partial #1}{\partial #2}}
\def\fd#1#2{\frac{\delta #1}{\delta #2}}
\def\wh#1{\widehat{#1}}
\def\bdot{\,\vb{\cdot}\,}
\def\btimes{\,\vb{\times}\,}

\def\bhat{\wh{{\sf b}}}
\def\cal#1{\mathcal{#1}}

\def\bhat{\wh{{\sf b}}}

\begin{document}

\title{Hamiltonian structure of a gauge-free gyrokinetic Vlasov-Maxwell model}

\author{Alain J.~Brizard$^{1,a)}$}
\affiliation{$^{1}$Department of Physics, Saint Michael's College, Colchester, VT 05439, USA \\ $^{a)}$Author to whom correspondence should be addressed: abrizard@smcvt.edu}

\begin{abstract}
The Hamiltonian structure of a set of gauge-free gyrokinetic Vlasov-Maxwell equations is presented in terms of a Hamiltonian functional and a gyrokinetic Vlasov-Maxwell bracket. The bracket is used to show that the gyrokinetic angular-momentum conservation law can be expressed in Hamiltonian form. The Jacobi property of the gyrokinetic Vlasov-Maxwell bracket is also demonstrated explicitly.
\end{abstract}

\date{\today}


\maketitle

\section{Introduction}

The Hamiltonian structure of several plasma physics models has been a topic of constant interest since the discovery of the Hamiltonian structures for the ideal magnetohydrodynamics \cite{Morrison_Greene_1980} and the Vlasov-Maxwell equations \cite{M,PJM_1982,MW,B}. The numerical algorithms derived from the Vlasov-Maxwell Hamiltonian structure were explored in several recent papers \cite{Squire_2012, Evstatiev_Shadwick_2013, He_2015, He_2016, GEMPIC_2017, Xiao_Qin_Liu_2018, Glasser_Qin_2020}. The generic guiding-center Vlasov-Maxwell bracket was presented by Morrison \cite{Morrison_2013}, while the generic gyrokinetic Vlasov-Maxwell bracket was presented by Burby {\it et al.} \cite{Burby_Brizard_2015, Burby_2015, Burby_2017,Burby_Tronci_2017}. The general Hamiltonian formulation for the reduced Vlasov-Maxwell equations was also derived by Lie-transform methods by Brizard {\it et al.} \cite{Brizard_2016}.

The guiding-center Vlasov-Maxwell equations suitable for Hamiltonian formulation were initially presented by Pfirsch and Morrison \cite{Pfirsch_Morrison_1985} and recently presented in simplified form (without guiding-center polarization) by Brizard and Tronci \cite{Brizard_Tronci_2016}, while two gauge-free gyrokinetic Vlasov-Maxwell models suitable for Hamiltonian formulation were presented by Burby and Brizard \cite{Burby_Brizard_2019} and Brizard \cite{Brizard_2020,Brizard_2021}. 

After having asymptotically eliminated the gyroangle $\zeta$ and constructed the gyroaction $J \equiv (mc/q)\,\mu$ as an adiabatic invariant ($\mu$ denotes the magnetic moment of a particle of mass $m$ and charge $q$), a reduced Lagrangian $L_{\rm g}$ is expressed in general form as
\begin{eqnarray}
L_{\rm g} &=& \left( \frac{q}{c}\,{\bf A} + \vb{\Pi}_{\rm g} \right) \bdot\frac{d{\bf X}}{dt} + J\;\frac{d\zeta}{dt} - \left( q\,\Phi \;+\frac{}{} K_{\rm g} \right),
 \label{eq:Lag_red}
 \end{eqnarray}
 where the reduced phase-space coordinates $Z^{\alpha} = ({\bf X},p_{\|},J,\zeta)$ include the reduced particle position ${\bf X}$ and the parallel kinetic momentum $p_{\|}$, the reduced symplectic momentum and kinetic energy are denoted 
 $\vb{\Pi}_{\rm g}$ and $K_{\rm g}$, respectively. We note that, because of the ``minimal-coupling'' potential terms $(q/c){\bf A}\bdot d{\bf X}/dt - q\,\Phi$, the reduced Lagrangian \eqref{eq:Lag_red} is invariant under an electromagnetic gauge transformation $(\Phi,{\bf A}) \rightarrow (\Phi - c^{-1}\partial\chi/\partial t, {\bf A} + \nabla\chi)$ since the reduced dynamics is invariant under the Lagrangian gauge transformation \cite{Brizard_2015} $L_{\rm g} \rightarrow L_{\rm g} + (q/c)\,d\chi/dt$, i.e., this transformation does not change the reduced equations of motion obtained from Eq.~\eqref{eq:Lag_red}. Hence, a gauge-free reduced Lagrangian formulation is obtained when the reduced symplectic momentum $\vb{\Pi}_{\rm g}$ and the reduced kinetic energy $K_{\rm g}$ in Eq.~\eqref{eq:Lag_red} depend on the electric and magnetic fields $({\bf E} \equiv -\nabla\Phi - c^{-1}\partial{\bf A}/\partial t,{\bf B} \equiv \nabla\btimes{\bf A})$ only. For example, in guiding-center Vlasov-Maxwell theory, Pfirsch and Morrison \cite{Pfirsch_Morrison_1985} used $\vb{\Pi}_{\rm gc} = p_{\|}\,\bhat + {\bf E}\btimes q\bhat/\Omega$ and $K_{\rm gc} = \mu\,B + |\vb{\Pi}_{\rm gc}|^{2}/2m$, while Brizard and Tronci \cite{Brizard_Tronci_2016} considered the simpler guiding-center Lagrangian with $\vb{\Pi}_{\rm gc} = p_{\|}\bhat$ and $K_{\rm gc} = \mu B + p_{\|}^{2}/2m$. The Hamiltonian structure for the Brizard-Tronci version of the guiding-center Vlasov-Maxwell equations was recently presented elsewhere \cite{Brizard_gcVM}, with an extensive proof of the Jacobi property for the guiding-center Vlasov-Maxwell bracket \cite{Brizard_proof_2021}.  
 
 In gyrokinetic Vlasov-Maxwell theory, on the other hand, the electromagnetic potentials in Eq.~\eqref{eq:Lag_red} are decomposed in terms of background and perturbed components: $(\Phi,{\bf A}) = (\epsilon\Phi_{1}, {\bf A}_{0} + \epsilon{\bf A}_{1})$, where 
 ${\bf B}_{0} = \nabla\btimes{\bf A}_{0}$ is the time-independent background (unperturbed) magnetic field, and the dimensionless parameter $\epsilon \ll 1$ orders the perturbation amplitudes in a manner consistent with standard gyrokinetic theory \cite{Brizard_Hahm_2007}. We note that, unless a gyrokinetic Vlasov-Maxwell model is formulated exclusively in terms of the perturbed electric and magnetic fields (with or without the minimal-coupling potential terms), the appearance of potentials in the gyrocenter kinetic energy $K_{\rm gy}$ prevents a Hamiltonian formulation. Hence, standard gyrokinetic Vlasov-Maxwell models \cite{Brizard_Hahm_2007} are not suitable for a Hamiltonian formulation. 
 
 The recent work of Burby and Brizard \cite{Burby_Brizard_2019} is suitable for a Hamiltonian formulation, however, since the gyrocenter symplectic momentum $\vb{\Pi}_{\rm gy} = p_{\|}\bhat_{0}$ is expressed only in terms of the unperturbed magnetic field, while the gyrocenter kinetic energy $K_{\rm gy}$ is a function of $({\bf E}_{1},{\bf B}_{1})$ up to second order in $\epsilon$ [see Eq.~\eqref{eq:K_gy} below]. The purpose of the present paper is, therefore, to derive the explicit Hamiltonian structure for the gauge-free gyrokinetic Vlasov-Maxwell equations presented by Burby and Brizard \cite{Burby_Brizard_2019}. This direct approach is in contrast to the formal derivation presented by Burby \cite{Burby_2015}. The case of the gauge-free gyrokinetic equations derived by Brizard \cite{Brizard_2020,Brizard_2021}, in which the gyrocenter symplectic momentum $\vb{\Pi}_{\rm gy} = p_{\|}\,\bhat_{0} + \epsilon\,\vb{\Pi}_{1{\rm gy}}({\bf E}_{1},{\bf B}_{1})$ includes first-order electromagnetic corrections, will be considered in future work.

The remainder of this paper is organized as follows. In Sec.~\ref{sec:gauge}, we present the gauge-free gyrokinetic Vlasov-Maxwell equations derived by Burby and Brizard \cite{Burby_Brizard_2019}, which are presented here in the drift-kinetic limit in order to simplify our presentation. In Sec.~\ref{sec:bracket}, the gyrokinetic Vlasov-Maxwell bracket is explicitly constructed from the Hamiltonian formulation of the gyrokinetic Vlasov-Maxwell equations. This gyrokinetic bracket structure is immediately applied to the proofs that the gyrokinetic entropy functional is a Casimir of the gyrokinetic Vlasov-Maxwell bracket and the gyrokinetic Vlasov-Maxwell toroidal angular momentum conservation law, first derived in variational (Lagrangian) form in Refs.~\cite{Hirvijoki_2020,Brizard_2021}, can be expressed in Hamiltonian form. In Sec.~\ref{sec:Jacobi}, the explicit proof of the Jacobi property of the gyrokinetic Vlasov-Maxwell bracket derived in Sec.~\ref{sec:bracket} is given, and a summary of our work is presented in Sec.~\ref{sec:summary}.

\section{\label{sec:gauge}Gauge-free Gyrokinetic Vlasov-Maxwell Equations}

We begin with the gauge-free gyrocenter single-particle Lagrangian 
\begin{eqnarray}
L_{\rm gy} &=& \left[ \frac{q}{c}\left({\bf A}_{0} + \epsilon\frac{}{}{\bf A}_{1}\right) + p_{\|}\,\bhat_{0} \;-\; J\,{\bf R}_{0}^{*}\right]\bdot\frac{d{\bf X}}{dt} + J\,\frac{d\zeta}{dt}\nonumber \\
 &&-\; \left(q\,\epsilon\,\Phi_{1} +\frac{}{} K_{\rm gy}\right) \equiv P_{\alpha}\,\frac{dZ^{\alpha}}{dt} - H_{\rm gy},
\label{eq:Lag_gy}
\end{eqnarray}
where the higher-order guiding-center corrections ${\bf R}_{0}^{*} \equiv {\bf R}_{0} + \frac{1}{2}\nabla\btimes\bhat_{0}$ include the background gyrogauge vector field ${\bf R}_{0}$ (which ensures that the guiding-center equations of motion are independent of the gyroangle as well as how the gyroangle is measured \cite{RGL_1983}), and the guiding-center polarization correction $\frac{1}{2}
\nabla\btimes\bhat_{0}$ \cite{Tronko_Brizard_2015}. Next, the gyrocenter kinetic energy is expanded up to second order in $\epsilon \ll 1$ \cite{Burby_Brizard_2019}:
\begin{eqnarray}
K_{\rm gy} &=& \frac{p_{\|}^{2}}{2m} \;+\; \mu \left( B_{0} \;+\; \epsilon\,B_{1\|} \;+\; \frac{\epsilon^{2}}{2B_{0}}\;|{\bf B}_{1}|^{2}\right) \nonumber \\
 &&-\,\epsilon\,\vb{\pi}_{\rm gc}\bdot\left( {\bf E}_{1} \;+\; \frac{p_{\|}\bhat_{0}}{mc}\btimes{\bf B}_{1}\right) \nonumber \\
  &&-\; \epsilon^{2}\;\frac{mc^{2}}{2B_{0}^{2}} \left|{\bf E}_{1} \;+\; (p_{\|}\bhat_{0}/mc)\btimes{\bf B}_{1}\right|^{2}
 \label{eq:K_gy}
 \end{eqnarray}
where $\vb{\pi}_{\rm gc}$ denotes the guiding-center electric-dipole moment \cite{Tronko_Brizard_2015}. For the sake of clarity, the gyrokinetic Vlasov-Maxwell model considered here is presented in its drift-kinetic limit, where finite-Larmor-radius (FLR) corrections are retained only through the guiding-center electric-dipole moment 
 $\vb{\pi}_{\rm gc}$. 

\subsection{Gyrocenter equations of motion}

The gyrocenter equations of motion are first derived from the gyrocenter Lagrangian \eqref{eq:Lag_gy} as Euler-Lagrange equations $\omega_{\alpha\beta}\,dZ^{\beta}/dt - \partial H_{\rm gy}/\partial Z^{\alpha}$:
\begin{eqnarray}
0 &=& \epsilon\,q{\bf E}_{1} \;-\; \nabla K_{\rm gy} \;+\; \frac{q}{c}\frac{d{\bf X}}{dt}\btimes{\bf B}^{*} \;-\; \frac{dp_{\|}}{dt}\,\bhat_{0}, \label{eq:EL_X} \\
0 &=& \bhat_{0}\bdot\frac{d{\bf X}}{dt} \;-\; \pd{K_{\rm gy}}{p_{\|}}, \label{eq:EL_p} \\
0 &=& -\,\frac{dJ}{dt} \;-\; \pd{K_{\rm gy}}{\zeta} \;\equiv\; -\,\frac{dJ}{dt}, \label{eq:EL_J} \\
0 &=& \frac{d\zeta}{dt} \;-\; \pd{K_{\rm gy}}{J} \;-\; {\bf R}_{0}^{*}\bdot\frac{d{\bf X}}{dt}, \label{eq:EL_zeta}
\end{eqnarray}
where $\omega_{\alpha\beta}({\bf X},p_{\|},\mu) \equiv \partial P_{\beta}/\partial Z^{\alpha} - \partial P_{\alpha}/\partial Z^{\beta}$. Equation \eqref{eq:EL_J} implies that the gyroaction $J$ (and the gyrocenter magnetic moment $\mu$) is a gyrocenter invariant as a result of the gyroangle-independence of the gyrocenter kinetic energy \eqref{eq:K_gy}, and the gyroangle $\zeta$ is an ignorable coordinate since Eq.~\eqref{eq:EL_zeta} is decoupled from the reduced gyrocenter equations of motion 
\eqref{eq:EL_X}-\eqref{eq:EL_p}, which are expressed in Hamiltonian form as
\begin{eqnarray}
\frac{d{\bf X}}{dt} &=& \left\{{\bf X},\; K_{\rm gy} \right\}_{\rm gy} \;+\; q\,\epsilon\,{\bf E}_{1}\bdot\{{\bf X},{\bf X}\}_{\rm gy}, \label{eq:X_dot_gy} \\
\frac{dp_{\|}}{dt} &=& \left\{p_{\|},\; K_{\rm gy} \right\}_{\rm gy} \;+\; q\,\epsilon\,{\bf E}_{1}\bdot\{{\bf X}, p_{\|}\}_{\rm gy}. \label{eq:p_dot_gy}  
 \end{eqnarray}
Here, the gyrocenter Poisson bracket
\begin{eqnarray}
\{ f,\; g\}_{\rm gy} &\equiv& \frac{{\bf B}^{*}}{B_{\|}^{*}}\bdot\left(\nabla f\;\pd{g}{p_{\|}} \;-\; \pd{f}{p_{\|}}\;\nabla g\right) \nonumber \\
 &&-\; \frac{c\bhat_{0}}{qB_{\|}^{*}}\bdot\nabla f\btimes\nabla g
\label{eq:PB_gy}
\end{eqnarray}
is used without the ignorable gyromotion canonical pair $(J,\zeta)$, with
\begin{equation}
\left. \begin{array}{rcl}
{\bf B}^{*} & \equiv & {\bf B}_{0}^{*} \;+\; \epsilon\,{\bf B}_{1} \\
B_{\|}^{*} & \equiv & \bhat_{0}\bdot{\bf B}^{*} \;=\; B_{\|0}^{*} \;+\; \epsilon\,B_{1\|} 
\end{array} \right\},
\label{eq:B*_def}
\end{equation}
where ${\bf B}_{0}^{*} = {\bf B}_{0} + (p_{\|} c/q)\,\nabla\btimes\bhat_{0} - (\mu mc^{2}/q^{2})\,\nabla\btimes{\bf R}_{0}^{*}$, which is gyrogauge invariant, and $B_{\|0}^{*} \equiv \bhat_{0}\bdot{\bf B}_{0}^{*}$. We note that the gyrocenter equations of motion \eqref{eq:X_dot_gy}-\eqref{eq:p_dot_gy} are gauge independent since they only involve the perturbed electromagnetic fields $({\bf E}_{1},{\bf B}_{1})$.

Next, we note that the gyrocenter Poisson bracket \eqref{eq:PB_gy} satisfies the Jacobi property for arbitrary functions $(f,g,h)$:
\begin{equation}
\left\{ \{f, g\}_{\rm gy},\frac{}{} h\right\}_{\rm gy} + \left\{ \{g, h\}_{\rm gy},\frac{}{} f\right\}_{\rm gy} + \left\{ \{h, f\}_{\rm gy},\frac{}{} g\right\}_{\rm gy} = 0,
\label{eq:gyPB_Jacobi}
\end{equation}
subject to the condition
\begin{equation}
\nabla\bdot{\bf B}^{*} \;=\; 0,
\label{eq:div_Bstar}
\end{equation}
which is satisfied by the definition \eqref{eq:B*_def}. We note that the gyrocenter Poisson bracket can be expressed in divergence form
\begin{equation}
\{f,\; g\}_{\rm gy} \;=\; \frac{1}{B_{\|}^{*}}\pd{}{Z^{\alpha}}\left( B_{\|}^{*}\frac{}{} f\;\left\{ Z^{\alpha},\; g\right\}_{\rm gy} \right),
\label{eq:PB_div}
\end{equation}
while the gyrocenter equations of motion \eqref{eq:X_dot_gy}-\eqref{eq:p_dot_gy} satisfy the gyrocenter Liouville equation
\begin{eqnarray}
\pd{B_{\|}^{*}}{t} & = & \bhat_{0}\bdot \epsilon\,\pd{{\bf B}_{1}}{t} \;=\; -\,c\,\bhat_{0}\bdot\nabla\btimes \epsilon\,{\bf E}_{1} \nonumber \\
 &=& -\,\nabla\bdot\left(B_{\|}^{*}\,\frac{d{\bf X}}{dt}\right) \;-\; \pd{}{p_{\|}}\left( B_{\|}^{*}\,\frac{dp_{\|}}{dt}\right).
\end{eqnarray}

\subsection{Gyrokinetic Vlasov-Maxwell equations}

With the help of the reduced gyrocenter equations of motion \eqref{eq:X_dot_gy}-\eqref{eq:p_dot_gy}, we now introduce the gyrokinetic Vlasov-Maxwell equations
\begin{eqnarray}
\pd{F_{\rm gy}}{t}  &=& -\,\nabla\bdot\left( F_{\rm gy}\;\frac{d{\bf X}}{dt}\right) \;-\; \pd{}{p_{\|}}\left( F_{\rm gy}\;\frac{dp_{\|}}{dt}\right), \label{eq:V_eq} \\
\pd{{\bf D}_{\rm gy}}{t} &=& c\,\nabla\btimes{\bf H}_{\rm gy} - 4\pi q\int_{P} F_{\rm gy}\,\frac{d{\bf X}}{dt}, \label{eq:Maxwell_eq} \\
\pd{{\bf B}_{1}}{t} &=& -\,c\,\nabla\btimes{\bf E}_{1}, \label{eq:Faraday_eq} 
\end{eqnarray}
where the gyrocenter phase-space density $F_{\rm gy} \equiv F\,B_{\|}^{*}$ is defined in terms of the gyrocenter Jacobian $B_{\|}^{*}$, so that the gyrokinetic Vlasov equation \eqref{eq:V_eq} is expressed in divergence form, the symbol $\int_{P}$ denotes an integration over $(p_{\|},\mu)$ in Eq.~\eqref{eq:Maxwell_eq}, and summation over particle species is implied throughout the work. 

The macroscopic gyrokinetic fields $({\bf D}_{\rm gy}, {\bf H}_{\rm gy})$ in Eq.~\eqref{eq:Maxwell_eq} are defined as
\begin{equation}
\left( \begin{array}{c}
{\bf D}_{\rm gy} \\
{\bf H}_{\rm gy}
\end{array} \right) \;=\; \left( \begin{array}{c}
\epsilon\,{\bf E}_{1} \;+\; 4\pi\,\mathbb{P}_{\rm gy} \\
{\bf B}_{0} + \epsilon\,{\bf B}_{1} \;-\; 4\pi\,\mathbb{M}_{\rm gy}
\end{array} \right),
\label{eq:DH_def}
\end{equation}
where the gyrocenter polarization and magnetization are defined in terms of the gyrocenter kinetic energy \eqref{eq:K_gy} as
\begin{eqnarray}
\mathbb{P}_{\rm gy} &=& -\;\epsilon^{-1}\int_{P} F_{\rm gy}\;\pd{K_{\rm gy}}{{\bf E}_{1}} \;\equiv\; \int_{P} F_{\rm gy}\;\vb{\pi}_{\rm gy}, \label{eq:P_gy} \\
 &=& \int_{P} F_{\rm gy} \left[ \vb{\pi}_{\rm gc} \;+\; \epsilon\,\frac{mc^{2}}{B_{0}^{2}}\left( {\bf E}_{1} \;+\; \frac{p_{\|}\bhat_{0}}{mc}\btimes{\bf B}_{1}\right) \right], \nonumber \\ 
\mathbb{M}_{\rm gy} &=& -\;\epsilon^{-1}\int_{P} F_{\rm gy}\;\pd{K_{\rm gy}}{{\bf B}_{1}} \label{eq:M_gy} \\
&=& \int_{P} F_{\rm gy} \left[ -\,\mu \left(\bhat_{0} \;+\; \epsilon\,\frac{{\bf B}_{1}}{B_{0}} \right) \;+\; \vb{\pi}_{\rm gy}\btimes\frac{p_{\|}\bhat_{0}}{mc} \right], \nonumber
\end{eqnarray}
which are expressed in their drift-kinetic dipole-moment form, whereas FLR corrections would involve higher-order multipole moments. 

The remaining Maxwell equations
\begin{equation}
\left. \begin{array}{rcl}
\nabla\bdot{\bf D}_{\rm gy} &=& 4\pi\,q\int_{P} F_{\rm gy} \\
\nabla\bdot{\bf B}_{1} &=& 0
\end{array} \right\}
\label{eq:div_DB}
\end{equation}
may be viewed as initial conditions for $({\bf D}_{\rm gy},{\bf B}_{1})$, since $\nabla\bdot(\partial{\bf D}_{\rm gy}/\partial t) = 4\pi\,\partial\varrho_{\rm gy}/\partial t = -\,4\pi\,\nabla\bdot{\bf J}_{\rm gy}$ follows from Eq.~\eqref{eq:Maxwell_eq}, which is a statement of the gyrokinetic charge conservation law, while $\nabla\bdot(\partial{\bf B}_{1}/\partial t) = 0$ follows from Eq.~\eqref{eq:Faraday_eq}.

\subsection{Hamiltonian gyrokinetic Vlasov-Maxwell equations}

In the Hamiltonian formulation of the gyrokinetic Vlasov-Maxwell equations \eqref{eq:V_eq}-\eqref{eq:Faraday_eq}, we begin with the gyrokinetic Hamiltonian functional \cite{Brizard_2021}
\begin{eqnarray}
{\cal H}_{\rm gy} &=& \int_{Z} F_{\rm gy}\;K_{\rm gy}({\bf E}_{1},{\bf B}_{1}) \;+\; \int_{X} \frac{\epsilon\,{\bf E}_{1}}{4\pi}\bdot{\bf D}_{\rm gy} \nonumber \\
 &&-\; \frac{1}{8\pi} \int_{X} \left( \epsilon^{2}\,|{\bf E}_{1}|^{2} \;-\; |{\bf B}_{0} + \epsilon\,{\bf B}_{1}|^{2} \right),
 \label{eq:Ham_gy}
 \end{eqnarray}
 which is derived by Noether method from a Lagrangian formulation of the gyrokinetic Vlasov-Maxwell equations \eqref{eq:V_eq}-\eqref{eq:Faraday_eq}. If we assume that ${\bf D}_{\rm gy}$ and ${\bf E}_{1}$ are functionally independent, we then find
 \begin{eqnarray}
 \fd{{\cal H}_{\rm gy}}{{\bf D}_{\rm gy}} &=& \epsilon\,{\bf E}_{1}/4\pi, \label{eq:H_D} \\
\epsilon^{-1} \fd{{\cal H}_{\rm gy}}{{\bf E}_{1}} &=& {\bf D}_{\rm gy}/4\pi \;-\; \epsilon\,{\bf E}_{1}/4\pi \;-\; \mathbb{P}_{\rm gy} \;=\; 0, \nonumber
\end{eqnarray}
where we used the definition \eqref{eq:P_gy} for the gyrocenter polarization. As can be seen from the definitions \eqref{eq:P_gy}-\eqref{eq:M_gy} of the gyrocenter polarization and magnetization, we might conclude that 
${\bf D}_{\rm gy} = {\bf D}_{\rm gy}[F_{\rm gy},{\bf E}_{1},{\bf B}_{1}]$ and ${\bf H}_{\rm gy} = {\bf H}_{\rm gy}[F_{\rm gy},{\bf E}_{1},{\bf B}_{1}]$ might be functionals of $(F_{\rm gy},{\bf E}_{1},{\bf B}_{1})$. The gyrokinetic Vlasov-Maxwell equations \eqref{eq:V_eq}-\eqref{eq:Faraday_eq}, however, clearly imply that the correct gyrokinetic fields are $(F_{\rm gy},{\bf D}_{\rm gy},{\bf B}_{1})$, with ${\bf E}_{1}[F_{\rm gy},{\bf D}_{\rm gy},{\bf B}_{1}]$ treated as a functional in Eq.~\eqref{eq:Ham_gy}. The reader is invited to consult Morrison's work \cite{Morrison_2013} and its application in gyrokinetic theory \cite{Burby_Brizard_2015} to learn how partial functional derivatives can be handled in terms of constitutive relations. 

In what follows, we will formulate a Hamiltonian representation of the gyrokinetic Vlasov-Maxwell equations in terms of the gyrokinetic fields $\vb{\Psi} = (F_{\rm gy},{\bf D}_{\rm gy},{\bf B}_{1})$. Hence, we shall also make use of the functional derivatives
\eqref{eq:H_D} and
\begin{eqnarray}
\fd{{\cal H}_{\rm gy}}{F_{\rm gy}} &=& K_{\rm gy}, \label{eq:H_F} \\
\epsilon^{-1}\fd{{\cal H}_{\rm gy}}{{\bf B}_{1}} &=& \int_{P} F_{\rm gy}\,\epsilon^{-1}\pd{K_{\rm gy}}{{\bf B}_{1}} \;+\; \frac{1}{4\pi}\left({\bf B}_{0} + \epsilon\,{\bf B}_{1}\right) \nonumber \\
 &=& {\bf H}_{\rm gy}/4\pi,
 \label{eq:H_B1}
 \end{eqnarray}
and we express the gyrokinetic Vlasov-Maxwell equations \eqref{eq:V_eq}-\eqref{eq:Faraday_eq} in Hamiltonian form
\begin{equation}
\pd{\Psi^{a}}{t} \;\equiv\; {\sf J}_{\rm gy}^{ab}(\vb{\Psi})\circ\fd{{\cal H}_{\rm gy}}{\Psi^{b}},
\end{equation}
where the gyrokinetic Vlasov-Maxwell Poisson operator ${\sf J}_{\rm gy}^{ab}(\vb{\Psi})\circ$ acts on functional derivatives of the gyrokinetic Hamiltonian functional \eqref{eq:Ham_gy}:
\begin{eqnarray}
\pd{F_{\rm gy}}{t} &=& -\;\pd{}{Z^{\alpha}}\left(F_{\rm gy}\;\left\{ Z^{\alpha},\; \fd{{\cal H}_{\rm gy}}{F_{\rm gy}} \right\}_{\rm gy} \right) \nonumber \\
 &&-\; \pd{}{Z^{\alpha}}\left(F_{\rm gy}\; 4\pi q\;\fd{{\cal H}_{\rm gy}}{{\bf D}_{\rm gy}}\bdot\{{\bf X},\; Z^{\alpha}\}_{\rm gy}\right) \nonumber \\
  &\equiv& {\sf J}_{\rm gy}^{Fb}(\vb{\Psi})\circ\fd{{\cal H}_{\rm gy}}{\Psi^{b}},
\label{eq:V_bracket} \\
 \pd{{\bf D}_{\rm gy}}{t} & = & 4\pi c\,\nabla\btimes\left( \epsilon^{-1}\fd{{\cal H}_{\rm gy}}{{\bf B}_{1}}\right) - 4\pi q \int_{P}F_{\rm gy}\;\frac{d{\bf X}}{dt} \nonumber \\
  &\equiv& {\sf J}_{\rm gy}^{{\bf D}b}(\vb{\Psi})\circ\fd{{\cal H}_{\rm gy}}{\Psi^{b}},
  \label{eq:B_bracket} \\
\pd{{\bf B}_{1}}{t} &=& -\,4\pi c\;\nabla\btimes\left(\epsilon^{-1}\fd{{\cal H}_{\rm gy}}{{\bf D}_{\rm gy}}\right) \nonumber \\
  &\equiv& {\sf J}_{\rm gy}^{{\bf B}b}(\vb{\Psi})\circ\fd{{\cal H}_{\rm gy}}{\Psi^{b}}.
  \label{eq:E_bracket}
  \end{eqnarray}
 The gyrokinetic Vlasov-Maxwell bracket will be constructed in the next Section from the gyrokinetic Vlasov-Maxwell equations \eqref{eq:V_bracket}-\eqref{eq:E_bracket} used in evaluating the time evolution of an arbitrary gyrokinetic functional
 ${\cal F}[F_{\rm gy},{\bf D}_{\rm gy},{\bf B}_{1}]$:
 \begin{eqnarray}
\pd{\cal F}{t} &=& \int_{Z}\pd{F_{\rm gy}}{t}\;\fd{\cal F}{F_{\rm gy}} + \int_{X}\left(\pd{{\bf D}_{\rm gy}}{t}\vb{\cdot}\fd{\cal F}{{\bf D}_{\rm gy}} + \pd{{\bf B}_{1}}{t}\vb{\cdot}\fd{\cal F}{{\bf B}_{1}}\right) \nonumber \\
 &\equiv& \left\langle \fd{\cal F}{\Psi^{a}}\left|\frac{}{}\right. {\sf J}_{\rm gy}^{ab}(\vb{\Psi})\circ\fd{{\cal H}_{\rm gy}}{\Psi^{b}} \right\rangle \;=\; \left[{\cal F},\frac{}{} {\cal H}_{\rm gy}\right]_{\rm gy},
\label{eq:F_fd}
\end{eqnarray} 
where the gyrokinetic Vlasov-Maxwell bracket is applied to functionals of the gyrokinetic fields $\vb{\Psi} = (F_{\rm gy},{\bf D}_{\rm gy},{\bf B}_{1})$ and integrations by parts may be performed. 

\section{\label{sec:bracket}Gyrokinetic Vlasov-Maxwell bracket}

In Eq.~\eqref{eq:F_fd}, the antisymmetric gyrokinetic Poisson operator ${\sf J}_{\rm gy}^{ab}(\vb{\Psi})\,\circ$ guarantees the antisymmetry property: $[{\cal F},{\cal G}]_{\rm gy} = -\,[{\cal G},{\cal F}]_{\rm gy}$; and the bilinearity of Eq.~\eqref{eq:F_fd} guarantees the Leibniz (product-rule) property: $[{\cal F},{\cal G}\,{\cal K}]_{\rm gy} = [{\cal F},{\cal G}]_{\rm gy}\,{\cal K} + {\cal G}\,[{\cal F},{\cal K}]_{\rm gy}$. The Jacobi property of the gyrokinetic Vlasov-Maxwell bracket is expressed as the requirement that the 
{\it Jacobiator}:
\begin{eqnarray}
{\cal Jac}[{\cal F},{\cal G},{\cal K}] &\equiv& \left[[{\cal F},{\cal G}]_{\rm gy},\frac{}{} {\cal K}\right]_{\rm gy} + \left[[{\cal G},{\cal K}]_{\rm gy},\frac{}{} {\cal F}\right]_{\rm gy} \nonumber \\
 &&+ \left[[{\cal K},{\cal F}]_{\rm gy},\frac{}{} {\cal G}\right]_{\rm gy} \;=\; 0
\label{eq:Jacobi}
\end{eqnarray} 
must vanish for arbitrary functionals $({\cal F},{\cal G},{\cal K})$, which imposes constraints on the Poisson operator ${\sf J}_{\rm gy}^{ab}(\vb{\Psi})$. 

 \begin{widetext}
From the gyrokinetic Vlasov-Maxwell equations \eqref{eq:V_bracket}-\eqref{eq:E_bracket}, we can now extract the gyrokinetic Vlasov-Maxwell bracket from Eq.~\eqref{eq:F_fd}, which is expressed in terms of two arbitrary gyrocenter functionals 
$({\cal F},{\cal G})$ as
 \begin{eqnarray}
 \left[{\cal F},\frac{}{}{\cal G}\right]_{\rm gy} &=& \int_{Z} F_{\rm gy} \left\{ \fd{{\cal F}}{F_{\rm gy}} ,\; \fd{\cal G}{F_{\rm gy}} \right\}_{\rm gy} + 4\pi q \int_{Z} F_{\rm gy}\;\left( \fd{\cal G}{{\bf D}_{\rm gy}}\bdot\left\{{\bf X},\; \fd{{\cal F}}{F_{\rm gy}} \right\}_{\rm gy} \;-\; \fd{\cal F}{{\bf D}_{\rm gy}}\bdot\left\{{\bf X},\;\fd{\cal G}{F_{\rm gy}} \right\}_{\rm gy} \right) 
   \label{eq:gyVM_bracket} \\
   &&+\;  (4\pi q)^{2} \int_{Z} F_{\rm gy} \left(\fd{\cal F}{{\bf D}_{\rm gy}}\vb{\cdot}\left\{{\bf X}, {\bf X}\right\}_{\rm gy}\vb{\cdot}\fd{\cal G}{{\bf D}_{\rm gy}}\right) + 4\pi c \int_{X} \left[ \fd{\cal F}{{\bf D}_{\rm gy}}\vb{\cdot}\nabla\vb{\times}\left(\epsilon^{-1}
   \fd{\cal G}{{\bf B}_{1}}\right) - \fd{\cal G}{{\bf D}_{\rm gy}}\vb{\cdot}\nabla\vb{\times}\left(\epsilon^{-1}\fd{\cal F}{{\bf B}_{1}}\right) \right].
\nonumber
 \end{eqnarray}
\end{widetext}
Here, the first term is the Vlasov sub-bracket, the next three terms (multiplied by first and second powers of $4\pi q$) represent the Interaction sub-bracket, and the last two terms (multiplied by $4\pi c$) represent the Maxwell sub-bracket. We note that the Interaction sub-bracket term proportional to $(4\pi q)^{2}$ does not appear in the standard Vlasov-Maxwell bracket \cite{M,PJM_1982,MW,B}, since the Poisson bracket $\{{\bf x},{\bf x}\} \equiv 0$ vanishes in particle phase space. The generic form of the gyrokinetic Vlasov-Maxwell bracket was first presented by Burby {\it et al.} \cite{Burby_Brizard_2015}. In the electrostatic limit, where ${\bf E}_{1} = -\,\nabla\Phi_{1}$ and ${\bf B}_{1} = 0$, the gyrokinetic Vlasov-Poisson bracket retains only the contributions from the Vlasov and Interaction sub-brackets. 

We postpone the proof of the Jacobi property \eqref{eq:Jacobi} until the next Section and, instead, we now look at two applications of the gyrokinetic Vlasov-Maxwell bracket \eqref{eq:gyVM_bracket}: first, we present the proof that the gyrokinetic entropy functional is a Casimir of the gyrokinetic bracket \eqref{eq:gyVM_bracket}; and, second, we present the proof that the gyrokinetic toroidal angular momentum conservation law can be expressed in Hamiltonian form.

\subsection{Gyrokinetic entropy functional}

A Casimir functional ${\cal C}$ associated with the gyrokinetic bracket \eqref{eq:gyVM_bracket} satisfies the equation $[{\cal C}, {\cal K}]_{\rm gy} = 0$, which holds for an arbitrary functional ${\cal K}$. It is well known that the gyrokinetic entropy functional
\begin{equation}
{\cal S}_{\rm gy}[F_{\rm gy},{\bf B}_{1}] \;\equiv\; -\;\int_{Z} F_{\rm gy}\;\ln(F_{\rm gy}/B_{\|}^{*})
\label{eq:S_gy}
\end{equation}
is a Casimir for the gyrokinetic bracket \eqref{eq:gyVM_bracket}, which is one example of the generic form ${\cal C}[F_{\rm gy},{\bf B}_{1}] = \int_{Z} B_{\|}^{*}\,C(F_{\rm gy}/B_{\|}^{*})$ \cite{Burby_Brizard_2015} for an arbitrary function $C(F)$, where $F = 
F_{\rm gy}/B_{\|}^{*}$.

From Eq.~\eqref{eq:S_gy}, using $\delta {\cal S}_{\rm gy}/\delta F_{\rm gy} = -\,1 - \ln F$ and $\epsilon^{-1}\delta{\cal S}_{\rm gy}/\delta{\bf B}_{1} = \int_{P} F\,\bhat_{0}$, we find
\begin{eqnarray}
\left[{\cal S}_{\rm gy},\frac{}{}{\cal K}\right]_{\rm gy} &=& -\;\int_{Z} B_{\|}^{*}\,\left\{ F, \fd{\cal K}{F_{\rm gy}}\right\}_{\rm gy} \nonumber \\
 &&-\; 4\pi q \int_{Z} \fd{\cal K}{{\bf D}_{\rm gy}}\bdot\left({\bf B}^{*}\;\pd{F}{p_{\|}} + \frac{c\bhat_{0}}{q}\btimes\nabla F\right) \nonumber \\
 &&-\; 4\pi c\int_{Z}\fd{\cal K}{{\bf D}_{\rm gy}}\bdot\nabla\times\left(F\,\bhat_{0}\right) \nonumber \\
  &=& 4\pi c\int_{Z}F\,\fd{\cal K}{{\bf D}_{\rm gy}}\bdot\left(\frac{q}{c}\pd{{\bf B}^{*}}{p_{\|}} - \nabla\btimes\bhat_{0}\right) \nonumber \\
   &=& 0,
 \end{eqnarray}
 where the first term vanishes since, according to Eq.~\eqref{eq:PB_div}, it is an exact phase-space divergence, while the remaining terms cancel out.
 
 The concept of gyrokinetic entropy can play an important role in the investigation of magnetized plasma turbulence (see Ref.~\cite{Schekochihin_2008}, for example). The gyrokinetic entropy functional \eqref{eq:S_gy} can also be used to formulate the gyrokinetic metriplectic evolution of an arbitrary gyrokinetic functional ${\cal F}$ \cite{Kaufman_1984,Morrison_1984,Morrison_1986}:
 \begin{equation}
 \pd{\cal F}{t} \;=\; \left[{\cal F},\frac{}{} {\cal H}_{\rm gy}\right]_{\rm gy} \;+\; \left({\cal F},\frac{}{} {\cal S}_{\rm gy}\right)_{\rm gy},
 \label{eq:gy_metri}
 \end{equation}
 in terms of a self-adjoint collisional bracket $(\;,\;)_{\rm gy}$ that conserves energy and momentum, i.e., $({\cal F},{\cal H}_{\rm gy})_{\rm gy} = 0$, and satisfies the second law of thermodynamics: $\partial{\cal S}_{\rm gy}/\partial t = ({\cal S}_{\rm gy},
 {\cal S}_{\rm gy})_{\rm gy} \geq 0$. This metriplectic formulation \cite{Kraus_Hirvijoki_2017,Hirvijoki_Burby_2020} can assist in the investigation of dissipative turbulent transport in magnetized plasmas based on structure-preserving algorithms.
 
\subsection{Gyrokinetic Vlasov-Maxwell angular momentum conservation law}

The conservation laws of energy-momentum and angular momentum for the gyrokinetic Vlasov-Maxwell \eqref{eq:V_bracket}-\eqref{eq:E_bracket} were recently derived by Brizard \cite{Brizard_2021} and Hirvijoki {\it et al.} \cite{Hirvijoki_2020}. As an application of the gyrocenter Vlasov-Maxwell bracket \eqref{eq:gyVM_bracket}, we explore the time evolution of the gyrokinetic Vlasov-Maxwell angular-momentum functional  \cite{Brizard_2021} 
\begin{eqnarray}
{\cal P}_{{\rm gy}\varphi}[F_{\rm gy},{\bf D}_{\rm gy},{\bf B}_{1}] &\equiv& \int_{Z} F_{\rm gy}\;P_{\varphi} \nonumber \\
 &&+ \int_{X} {\bf D}_{\rm gy}\btimes\frac{\epsilon\,{\bf B}_{1}}{4\pi\,c}\bdot\pd{\bf X}{\varphi},
\label{eq:P_gyVM}
\end{eqnarray}
where the gyrocenter toroidal angular momentum
\begin{eqnarray}
P_{\varphi} &=& \left[\frac{q}{c}{\bf A}_{0} \;+\; p_{\|}\,\bhat_{0} \;-\; (mc/q)\,\mu\,{\bf R}_{0}^{*}\right]\bdot\pd{\bf X}{\varphi} \nonumber \\
 &\equiv& \frac{q}{c}\,{\bf A}_{0}^{*}\bdot\pd{\bf X}{\varphi}
\label{eq:P_phi}
\end{eqnarray}
includes higher-order guiding-center corrections \cite{Tronko_Brizard_2015}.

We now evaluate the Hamiltonian evolution of the gyrokinetic functional \eqref{eq:P_gyVM}:
 \begin{eqnarray}
 \pd{{\cal P}_{{\rm gy}\varphi}}{t} &=& \left[ {\cal P}_{{\rm gy}\varphi},\frac{}{} {\cal H}_{\rm gy} \right]_{\rm gy} \nonumber \\
 &=& \int_{Z} F_{\rm gy} \left(\frac{dP_{\varphi}}{dt} \;-\; \frac{q}{c}\frac{d{\bf X}}{dt}\btimes \epsilon{\bf B}_{1}\bdot\pd{\bf X}{\varphi} \right)
\nonumber \\
 &&+\; \int_{X} \left(\frac{\epsilon {\bf B}_{1}}{4\pi}\btimes\pd{\bf X}{\varphi}\right)\bdot\nabla\btimes{\bf H}_{\rm gy} \nonumber \\
  &&-\; \int_{X} \frac{\epsilon {\bf E}_{1}}{4\pi}\bdot\nabla\btimes\left(\pd{\bf X}{\varphi}\btimes{\bf D}_{\rm gy}\right),
  \label{eq:gy_P_phi} 
 \end{eqnarray}
where the functional derivatives of the gyrocenter Hamiltonian functional \eqref{eq:Ham_gy} are given in Eqs.~\eqref{eq:H_D}-\eqref{eq:H_B1}, and the functional derivatives of the gyrocenter Vlasov-Maxwell angular momentum \eqref{eq:P_gyVM} are
\begin{equation}
\left( \begin{array}{c}
\delta{\cal P}_{{\rm gy}\varphi}/\delta F_{\rm gy} \\
4\pi c\,\delta{\cal P}_{{\rm gy}\varphi}/\delta{\bf D}_{\rm gy} \\
4\pi c\,\delta{\cal P}_{{\rm gy}\varphi}/\delta(\epsilon\,{\bf B}_{1})
\end{array} \right) = \left( \begin{array}{c}
P_{\varphi}  \\
\epsilon\,{\bf B}_{1}\btimes\partial{\bf X}/\partial\varphi \\
(\partial{\bf X}/\partial\varphi)\btimes{\bf D}_{\rm gy}
\end{array} \right).
\label{eq:delta_Pgy}
\end{equation}
If we ignore exact spatial derivatives (which vanish when integrated over space), the second term in Eq.~\eqref{eq:gy_P_phi} yields the non-vanishing terms
\begin{eqnarray}
 &&\left(\frac{\epsilon {\bf B}_{1}}{4\pi}\btimes\pd{\bf X}{\varphi}\right)\bdot\nabla\btimes{\bf H}_{\rm gy} \nonumber \\
  &=& {\bf H}_{\rm gy}\bdot\frac{\epsilon}{4\pi}\pd{{\bf B}_{1}}{\varphi} \;-\; \frac{\epsilon{\bf B}_{1}}{4\pi}\bdot\nabla\left(\pd{\bf X}{\varphi}\right)\bdot{\bf H}_{\rm gy} \nonumber \\
   &=& -\,\frac{\epsilon{\bf B}_{1}}{4\pi}\bdot\left( \pd{{\bf B}_{0}}{\varphi} - \wh{\sf z}\btimes{\bf B}_{0}\right) - \mathbb{M}_{\rm gy}\bdot\epsilon\left(\pd{{\bf B}_{1}}{\varphi} - \wh{\sf z}\btimes{\bf B}_{1}\right) \nonumber \\
  &=& \int_{P} F_{\rm gy}\;\pd{K_{\rm gy}}{{\bf B}_{1}}\bdot\pd{{\bf B}_{1}}{\varphi}  \;+\; \wh{\sf z}\bdot\left(\epsilon\,{\bf B}_{1}\btimes\mathbb{M}_{\rm gy}\right),
  \label{eq:H_eq}
 \end{eqnarray}
 where we used the definitions \eqref{eq:DH_def} and \eqref{eq:M_gy} for the gyrokinetic H-field and the gyrocenter magnetization, respectively, and we used the axisymmetric vector identity 
 \begin{equation}
 \partial{\bf B}_{0}/\partial\varphi \;=\; \wh{\sf z}\btimes{\bf B}_{0},
 \label{eq:B0_phi}
 \end{equation}
 and the vector identity
 \begin{equation}
 {\bf V}{\bf W}:\nabla(\partial{\bf X}/\partial\varphi) \;=\; \wh{\sf z}\bdot({\bf W}\btimes{\bf V}), 
 \label{eq:vector_id}
 \end{equation}
 which holds for arbitrary vectors fields $({\bf V},{\bf W})$. Next, the third term in Eq.~\eqref{eq:gy_P_phi} yields the non-vanishing terms
 \begin{eqnarray}
 &&-\, \frac{\epsilon {\bf E}_{1}}{4\pi}\bdot\nabla\btimes\left(\pd{\bf X}{\varphi}\btimes{\bf D}_{\rm gy}\right) \nonumber \\
  &=& -\,\frac{\epsilon {\bf E}_{1}}{4\pi}\bdot\pd{\bf X}{\varphi}\;(\nabla\bdot{\bf D}_{\rm gy}) - {\bf D}_{\rm gy}\bdot\frac{\epsilon}{4\pi}\pd{{\bf E}_{1}}{\varphi} \nonumber \\
   &&-\; {\bf D}_{\rm gy}\bdot\nabla\left(\pd{\bf X}{\varphi}\right)\bdot\frac{\epsilon{\bf E}_{1}}{4\pi} \nonumber \\
  &=& \int_{P} F_{\rm gy}\;\left(\pd{K_{\rm gy}}{{\bf E}_{1}}\bdot\pd{{\bf E}_{1}}{\varphi}  - \epsilon\,q{\bf E}_{1}\bdot\pd{\bf X}{\varphi} \right) \nonumber \\
   &&+\; \wh{\sf z}\bdot\left(\epsilon\,{\bf E}_{1}\btimes\mathbb{P}_{\rm gy}\right), 
\label{eq:D_eq}
\end{eqnarray}
where we used the vector identity \eqref{eq:vector_id} and the definition \eqref{eq:P_gy} for the gyrocenter polarization, as well as the gyrokinetic Poisson equation in Eq.~\eqref{eq:div_DB}. Hence, by combining Eqs.~\eqref{eq:H_eq} and \eqref{eq:D_eq}, Eq.~\eqref{eq:gy_P_phi} becomes
\begin{eqnarray}
 \pd{{\cal P}_{{\rm gy}\varphi}}{t} &=& \int_{Z} F_{\rm gy} \left[\frac{dP_{\varphi}}{dt} - \epsilon\,q \left({\bf E}_{1} + \frac{1}{c}\frac{d{\bf X}}{dt}\vb{\times}{\bf B}_{1}\right)\vb{\cdot}\pd{\bf X}{\varphi} \right] \nonumber \\
  &&+\; \int_{Z} F_{\rm gy} \left(\pd{K_{\rm gy}}{{\bf E}_{1}}\bdot\pd{{\bf E}_{1}}{\varphi} + \pd{K_{\rm gy}}{{\bf B}_{1}}\bdot\pd{{\bf B}_{1}}{\varphi}\right) \nonumber \\
   &&+\; \int_{X} \wh{\sf z}\bdot\left( \epsilon\,{\bf E}_{1}\btimes\mathbb{P}_{\rm gy} \;+\frac{}{} \epsilon\,{\bf B}_{1}\btimes\mathbb{M}_{\rm gy} \right). 
   \label{eq:Pdot_2}
 \end{eqnarray}
 Next, using the explicit expressions for the gyrocenter polarization and magnetization \eqref{eq:P_gy}-\eqref{eq:M_gy}, the last terms in Eq.~\eqref{eq:Pdot_2} become the polarization and magnetization torques
 \begin{eqnarray}
 && \int_{X} \wh{\sf z}\bdot\left( \epsilon\,{\bf E}_{1}\btimes\mathbb{P}_{\rm gy} \;+\frac{}{} \epsilon\,{\bf B}_{1}\btimes\mathbb{M}_{\rm gy} \right) \nonumber \\
  &&=\; \int_{Z} F_{\rm gy} \;\wh{\sf z}\bdot\left(\mu\,\bhat_{0}\btimes\epsilon {\bf B}_{1} \;+\; \epsilon{\bf E}_{1}\btimes\vb{\pi}_{\rm gy} \right) \nonumber \\
   &&+\; \int_{Z} F_{\rm gy} \;\wh{\sf z}\bdot\left[ \epsilon{\bf B}_{1}\btimes\left(\vb{\pi}_{\rm gy}\btimes\frac{p_{\|}\bhat_{0}}{mc}\right) \right].
  \end{eqnarray}
  
  We now write the full expression for $\partial K_{\rm gy}/\partial\varphi$:
  \begin{equation}
  \pd{K_{\rm gy}}{\varphi} \;=\; \frac{\partial^{\prime}K_{\rm gy}}{\partial\varphi} +  \left(\pd{K_{\rm gy}}{{\bf E}_{1}}\bdot\pd{{\bf E}_{1}}{\varphi} + \pd{K_{\rm gy}}{{\bf B}_{1}}\bdot\pd{{\bf B}_{1}}{\varphi}\right),
  \label{eq:K_phi}
  \end{equation}
  where $\partial^{\prime}K_{\rm gy}/\partial\varphi$ denotes the derivative of the gyrocenter kinetic energy \eqref{eq:K_gy} at constant perturbed fields $({\bf E}_{1},{\bf B}_{1})$:
  \begin{eqnarray}
  \frac{\partial^{\prime}K_{\rm gy}}{\partial\varphi} &=& \wh{\sf z}\bdot\left(\mu\,\bhat_{0}\btimes\epsilon {\bf B}_{1} \;+\; \epsilon{\bf E}_{1}\btimes\vb{\pi}_{\rm gy} \right) \nonumber \\
   &&-\; \wh{\sf z}\bdot\left[ \frac{p_{\|}\bhat_{0}}{mc}\btimes\left( \epsilon{\bf B}_{1}\btimes\vb{\pi}_{\rm gy}\right)\right] \nonumber \\
    &&-\; \wh{\sf z}\bdot\left[ \vb{\pi}_{\rm gy}\btimes\left( \frac{p_{\|}\bhat_{0}}{mc}\btimes \epsilon{\bf B}_{1}\right) \right],
   \end{eqnarray}
  where we used Eq.~\eqref{eq:B0_phi}. If we now combine these expressions in Eq.~\eqref{eq:Pdot_2}, we find
\begin{eqnarray}
 \pd{{\cal P}_{{\rm gy}\varphi}}{t} &=& \int_{Z} F_{\rm gy} \left[\frac{dP_{\varphi}}{dt} - \pd{K_{\rm gy}}{\varphi} \right. \nonumber \\
  &&\left.-\; \epsilon\,q \left({\bf E}_{1} + \frac{1}{c}\frac{d{\bf X}}{dt}\vb{\times} {\bf B}_{1}\right)\vb{\cdot}\pd{\bf X}{\varphi} \right],
 \label{eq:P_dot_final}
 \end{eqnarray} 
 after using the vector identity 
 \[ {\bf U}\btimes({\bf V}\btimes{\bf W}) \;+\; {\bf V}\btimes({\bf W}\btimes{\bf U}) \;+\; {\bf W}\btimes({\bf U}\btimes{\bf V}) \;=\; 0, \]
 with ${\bf U} = \epsilon {\bf B}_{1}$, ${\bf V} = \vb{\pi}_{\rm gy}$, and ${\bf W} = p_{\|}\bhat_{0}/mc$. Lastly, using the equation of motion \eqref{eq:P_dot} for the gyrocenter azimuthal angular momentum, we arrive at the conservation law
 \begin{equation}
 \pd{{\cal P}_{{\rm gy}\varphi}}{t} \;=\; \left[ {\cal P}_{{\rm gy}\varphi},\frac{}{} {\cal H}_{\rm gy}\right]_{\rm gy} \;=\; 0.
 \label{eq:P_phi_Ham}
 \end{equation}
 This conservation law was derived by Noether method \cite{Brizard_2021} in the variational (Lagrangian) formulation of the gyrokinetic Vlasov-Maxwell equations \eqref{eq:V_eq}-\eqref{eq:Faraday_eq}. Here, we have shown that this conservation law can also be expressed in Hamiltonian form with the help of the gyrokinetic Vlasov-Maxwell bracket \eqref{eq:gyVM_bracket}.
 
\section{\label{sec:Jacobi}Jacobi Property of the Gyrokinetic Vlasov-Maxwell Bracket}

We now verify that the gyrokinetic bracket \eqref{eq:gyVM_bracket} satisfies the Jacobi property \eqref{eq:Jacobi}. The proof will rely on several Poisson-bracket identities derived from the gyrocenter Poisson bracket \eqref{eq:PB_gy}.

According to the Bracket theorem \cite{PJM_1982,Morrison_2013},  the proof of the Jacobi property involves only the explicit dependence of the gyrocenter Vlasov-Maxwell bracket \eqref{eq:gyVM_bracket}, where the gyrokinetic  Vlasov-Maxwell Poisson operator ${\sf J}_{\rm gy}^{ab}(F_{\rm gy}, {\bf B}_{1})\circ$ is independent of the gyrokinetic displacement field ${\bf D}_{\rm gy}$, where we note that the dependence on the magnetic field ${\bf B}_{1}$ enters through Eq.~\eqref{eq:B*_def} appearing in the gyrocenter Poisson bracket \eqref{eq:PB_gy}. Hence, we can write the double-bracket involving three arbitrary gyrocenter functionals $({\cal F}, {\cal G}, {\cal K})$:
\begin{eqnarray}
\left[[{\cal F},{\cal G}]_{\rm gy},\frac{}{} {\cal K}\right]_{\rm gy}^{P} &=& \int_{Z} F_{\rm gy} \left\{ \frac{\delta^{P}[{\cal F}, {\cal G}]_{\rm gy}}{\delta F_{\rm gy}}, \fd{\cal K}{F_{\rm gy}} \right\}_{\rm gy} \nonumber \\
 &&+ 4\pi q \int_{Z} F_{\rm gy}\,\fd{\cal K}{{\bf D}_{\rm gy}}\vb{\cdot}\left\{ {\bf X},  \frac{\delta^{P}[{\cal F}, {\cal G}]_{\rm gy}}{\delta F_{\rm gy}} \right\}_{\rm gy} \nonumber \\
 &&- 4\pi c\int_{X}  \frac{\delta^{P}[{\cal F}, {\cal G}]_{\rm gy}}{\epsilon\delta {\bf B}_{1}}\bdot\nabla\btimes\fd{\cal K}{{\bf D}_{\rm gy}},
 \label{eq:Jac_fg-k}
\end{eqnarray}
where the terms involving $\delta^{P}[{\cal F}, {\cal G}]_{\rm gy}/\delta{\bf D}_{\rm gy}$ vanish on the basis of the Bracket theorem. 

Here, the Poisson variation $\delta^{P}$ of the bracket \eqref{eq:gyVM_bracket} only involves variations with respect to $(F_{\rm gy}, {\bf B}_{1})$:
\begin{widetext}
\begin{eqnarray}
\delta^{P}[{\cal F},{\cal G}]_{\rm gy} &=& \int_{Z} \left( \delta F_{\rm gy} - \frac{F_{\rm gy}}{B_{\|}^{*}}\,\bhat_{0}\bdot\epsilon\,\delta{\bf B}_{1}\right) \left[\{f,\; g\}_{\rm gy} \;+\frac{}{} 4\pi q\,\left({\bf G}\bdot\{{\bf X},f\}_{\rm gy} \;-\frac{}{} 
{\bf F}\bdot\{{\bf X},g\}_{\rm gy} \right) + (4\pi q)^{2}{\bf F}\vb{\cdot}\{{\bf X},{\bf X}\}_{\rm gy}\vb{\cdot}{\bf G}\right] \nonumber \\
 &&+\; \int_{Z} F_{\rm gy}\,\frac{\epsilon\,\delta{\bf B}_{1}}{B_{\|}^{*}}\bdot\left[ \left(\nabla f \;-\frac{}{} 4\pi q\,{\bf F}\right)\pd{g}{p_{\|}} \;-\; \left(\nabla g \;-\frac{}{} 4\pi q\,{\bf G}\right)\pd{f}{p_{\|}} \right],
 \label{eq:delta_Poisson}
\end{eqnarray}
where we use the notation $(f,g,k) \equiv (\delta{\cal F}/\delta F_{\rm gy}, \delta{\cal G}/\delta F_{\rm gy}, \delta{\cal K}/\delta F_{\rm gy})$ and $({\bf F},{\bf G},{\bf K}) \equiv (\delta{\cal F}/\delta {\bf D}_{\rm gy}, \delta{\cal G}/\delta {\bf D}_{\rm gy}, 
\delta{\cal K}/\delta {\bf D}_{\rm gy})$. In addition, $\delta B_{\|}^{*} = \bhat_{0}\bdot\epsilon\delta{\bf B}_{1}$ represents the variation of the Jacobian, which appears in the denominator of the gyrocenter Poisson bracket \eqref{eq:PB_gy}, while $\delta{\bf B}^{*} = \epsilon\,\delta{\bf B}_{1}$ appears only in the first term of Eq.~\eqref{eq:PB_gy}. In the first two terms of Eq.~\eqref{eq:Jac_fg-k}, we therefore have the Vlasov sub-bracket
\begin{eqnarray}
\left\{ \frac{\delta^{P}[{\cal F}, {\cal G}]_{\rm gy}}{\delta F_{\rm gy}}, \fd{\cal K}{F_{\rm gy}} \right\}_{\rm gy} &=& \left\{ \{f,\; g\}_{\rm gy},\frac{}{} k\right\}_{\rm gy} \;+\; 4\pi q \left\{ \left({\bf G}\bdot\{{\bf X},f\}_{\rm gy} \;-\frac{}{} {\bf F}\bdot\{{\bf X},g\}_{\rm gy}\right),\frac{}{} k\right\}_{\rm gy} \nonumber \\
 &&+\; (4\pi q)^{2} \left\{ {\bf F}\bdot\{{\bf X},{\bf X}\}_{\rm gy}\bdot{\bf G},\frac{}{} k\right\}_{\rm gy}, 
 \label{eq:V}
 \end{eqnarray}
 and the Interaction sub-bracket
 \begin{eqnarray}
4\pi q \fd{\cal K}{{\bf D}_{\rm gy}}\bdot\left\{ {\bf X},  \frac{\delta^{P}[{\cal F}, {\cal G}]_{\rm gy}}{\delta F_{\rm gy}} \right\}_{\rm gy} &=& 4\pi q\,{\bf K}\bdot\left\{{\bf X},\frac{}{} \{f,\; g\}_{\rm gy}\right\}_{\rm gy} + (4\pi q)^{2}\;{\bf K}\bdot\left\{{\bf X},\frac{}{}
\left({\bf G}\bdot\{{\bf X},f\}_{\rm gy} \;-\frac{}{} {\bf F}\bdot\{{\bf X},g\}_{\rm gy}\right)\right\}_{\rm gy} \nonumber \\
 &&+\; (4\pi q)^{3}\; {\bf K}\bdot\left\{ {\bf X},\frac{}{} {\bf F}\bdot\{{\bf X},{\bf X}\}_{\rm gy}\bdot{\bf G}\right\}_{\rm gy},
 \label{eq:I}
 \end{eqnarray}
while the third term in Eq.~\eqref{eq:Jac_fg-k} is the Maxwell sub-bracket
 \begin{eqnarray}
 - 4\pi c\int_{X}  \frac{\delta^{P}[{\cal F}, {\cal G}]_{\rm gy}}{\epsilon\,\delta {\bf B}_{1}}\bdot\nabla\btimes\fd{\cal K}{{\bf D}_{\rm gy}} &=& -\,4\pi q\int_{Z} \frac{cF_{\rm gy}}{qB_{\|}^{*}}\;\nabla\btimes{\bf K}\bdot\left[ \left(\nabla f \;-\frac{}{} 4\pi q\,{\bf F}\right)\pd{g}{p_{\|}} \;-\; \left(\nabla g \;-\frac{}{} 4\pi q\,{\bf G}\right)\pd{f}{p_{\|}} \right] \nonumber \\
  &&+\; 4\pi q\int_{Z} F_{\rm gy}\;\frac{c\bhat_{0}}{qB_{\|}^{*}}\bdot\nabla\btimes{\bf K} \left[\{f,\; g\}_{\rm gy} \;+\frac{}{} 4\pi q\,\left({\bf G}\bdot\{{\bf X},f\}_{\rm gy} \;-\frac{}{} {\bf F}\bdot\{{\bf X},g\}_{\rm gy}\right) \right] \nonumber \\
   &&+\;  (4\pi q)^{3}\int_{Z} F_{\rm gy}\;\frac{c\bhat_{0}}{qB_{\|}^{*}}\bdot\nabla\btimes{\bf K} \left({\bf F}\bdot\{{\bf X},\frac{}{}{\bf X}\}_{\rm gy}\bdot{\bf G}\right).
  \label{eq:M}
 \end{eqnarray}
\end{widetext}
From these terms, it is clear that the proof of the Jacobi property \eqref{eq:Jacobi} must hold separately for each power of $4\pi q$:
\begin{equation}
{\cal Jac}[{\cal F},{\cal G},{\cal K}] \;\equiv\; \sum_{n=0}^{3} (4\pi q)^{n}\int_{Z} F_{\rm gy} \;{\sf Jac}_{n}[{\cal F},{\cal G},{\cal K}] ,
\label{eq:Jacobi_0123}
\end{equation}
where each Jacobiator term ${\sf Jac}_{n}[{\cal F},{\cal G},{\cal K}]$ involves the gyrocenter Poisson bracket \eqref{eq:PB_gy}. At zeroth order, for example, the Vlasov sub-bracket \eqref{eq:V} yields
\begin{eqnarray}
{\sf Jac}_{0}[{\cal F},{\cal G},{\cal K}] &=& \left\{ \{f, g\}_{\rm gy},\frac{}{} k\right\}_{\rm gy} \;+\; \left\{ \{g, k\}_{\rm gy},\frac{}{} f\right\}_{\rm gy} \nonumber \\
 &&+\; \left\{ \{k, f\}_{\rm gy},\frac{}{} g\right\}_{\rm gy} \;=\; 0,
\label{eq:Jacobi_0}
\end{eqnarray}
which vanishes because of the Jacobi property \eqref{eq:gyPB_Jacobi} of the gyrocenter Poisson bracket \eqref{eq:PB_gy}. In what follows, we will use cyclic permutations (denoted by $\leftturn$) of the functionals $({\cal F},{\cal G},{\cal K})$ in order to combine similar terms from the sub-brackets \eqref{eq:V}-\eqref{eq:M} at the next three orders in $4\pi q$.

\subsection{First-order Jacobi property}

By using the Leibniz property of the gyrocenter Poisson bracket \eqref{eq:PB_gy}, the cyclic permutations of the Vlasov sub-bracket \eqref{eq:V} include the first-order terms
\begin{eqnarray}
 &&\{{\bf K}, f\}_{\rm gy}\bdot\{{\bf X}, g\}_{\rm gy} \;-\; \{{\bf K}, g\}_{\rm gy}\bdot\{{\bf X}, f\}_{\rm gy} \nonumber \\
  &&+\; {\bf K}\bdot\left( \left\{ \{{\bf X},g\}_{\rm gy},\frac{}{} f \right\}_{\rm gy} + \left\{ \{f, {\bf X}\}_{\rm gy},\frac{}{} g \right\}_{\rm gy}\right),
\label{eq:V_1}
\end{eqnarray}
while the cyclic permutations of the Interaction sub-bracket \eqref{eq:I} include
\begin{equation}
{\bf K}\bdot\left\{{\bf X},\frac{}{} \{f,\; g\}_{\rm gy}\right\}_{\rm gy} \;=\; {\bf K}\bdot\left\{\{g,\;f\}_{\rm gy},\frac{}{} {\bf X}\right\}_{\rm gy},
\label{eq:I_1}
\end{equation}
where we used the antisymmetry of the gyrocenter Poisson bracket. Lastly, the first-order terms in cyclic permutations of the Maxwell sub-bracket \eqref{eq:M} are
\begin{equation}
\frac{c\bhat_{0}}{qB_{\|}^{*}}\vb{\cdot}\nabla\vb{\times}{\bf K}\;\{f,\; g\}_{\rm gy} - \frac{c}{qB_{\|}^{*}}\nabla\vb{\times}{\bf K}\vb{\cdot}\left(\nabla f\,\pd{g}{p_{\|}} - \nabla g\,\pd{f}{p_{\|}}\right).
\label{eq:M_11}
\end{equation}
Using the gyrocenter Poisson bracket identity
\begin{eqnarray*}
\frac{c}{qB_{\|}^{*}}\left(\nabla f\,\pd{g}{p_{\|}} - \nabla g\,\pd{f}{p_{\|}}\right) &=& \frac{c\bhat_{0}}{qB_{\|}^{*}}\,\{f,\; g\}_{\rm gy} \\
 &&+\; \{{\bf X},\; f\}_{\rm gy}\btimes\{{\bf X},\;g\}_{\rm gy}
 \end{eqnarray*}
 Eq.~\eqref{eq:M_11} becomes
 \begin{eqnarray}
  &&\nabla\btimes{\bf K}\bdot\{{\bf X},\; g\}_{\rm gy}\btimes\{{\bf X},\;f\}_{\rm gy} \nonumber \\
   &&=\; \{{\bf K}, g\}_{\rm gy}\bdot\{{\bf X}, f\}_{\rm gy} \;-\; \{{\bf K}, f\}_{\rm gy}\bdot\{{\bf X}, g\}_{\rm gy},
  \label{eq:M_1}
\end{eqnarray}
where we used the identity 
\begin{equation}
\{{\bf U}, v\}_{\rm gy} \equiv \{{\bf X},v\}_{\rm gy}\bdot\nabla{\bf U}
\label{eq:Uv_id}
\end{equation}
for a vector field ${\bf U}$ assumed to be independent of $p_{\|}$, while $v({\bf X},p_{\|})$ is an arbitrary gyrocenter function.

By combining Eqs.~\eqref{eq:V_1}, \eqref{eq:I_1}, and \eqref{eq:M_1}, we obtain the first-order term in the Jacobiator \eqref{eq:Jacobi_0123}:
\begin{eqnarray}
{\sf Jac}_{1}[{\cal F},{\cal G},{\cal K}] &=& K_{i}\left[ \left\{ \{X^{i},g\}_{\rm gy},\frac{}{} f \right\}_{\rm gy} \frac{}{}\right. \nonumber \\
 &&+\; \left\{ \{f, X^{i}\}_{\rm gy},\frac{}{} g \right\}_{\rm gy} \nonumber \\
 &&\left.+\frac{}{} \left\{\{g,\;f\}_{\rm gy},\frac{}{} X^{i}\right\}_{\rm gy}\right] + \leftturn = 0,
 \label{eq:Jacobi_1}
\end{eqnarray}
which vanishes because of the Jacobi property \eqref{eq:gyPB_Jacobi} of the gyrocenter Poisson bracket \eqref{eq:PB_gy}.

\subsection{Second-order Jacobi property}

Cyclic permutations of the Vlasov sub-bracket \eqref{eq:V} include the second-order terms
\begin{eqnarray}
 \left\{ {\bf K}\vb{\cdot}\{{\bf X},{\bf X}\}_{\rm gy}\vb{\cdot}{\bf F},\frac{}{} g\right\}_{\rm gy} &=& F_{i}\,K_{j} \left\{ \{X^{j}, X^{i}\}_{\rm gy},\frac{}{} g\right\}_{\rm gy} \nonumber \\
  &&+\; \{{\bf K}, g\}_{\rm gy}\vb{\cdot}\{{\bf X},{\bf X}\}_{\rm gy}\vb{\cdot}{\bf F} \nonumber \\
   &&+\; {\bf K}\vb{\cdot}\{{\bf X},{\bf X}\}_{\rm gy}\vb{\cdot}\{{\bf F}, g\}_{\rm gy},
   \label{eq:V2}
 \end{eqnarray}
where summation over repeated indices is implied in the first term.  Next, using the identity \eqref{eq:Uv_id}, the second and third terms become
\begin{equation}
\{{\bf X},g\}_{\rm gy}\bdot\left( \nabla{\bf K}\bdot\frac{c\bhat_{0}}{qB_{\|}^{*}}\btimes{\bf F} - \nabla{\bf F}\bdot\frac{c\bhat_{0}}{qB_{\|}^{*}}\btimes{\bf K}\right),
\label{eq:V2_id}
\end{equation}
where we used the Poisson-bracket identity 
\begin{equation}
{\bf U}\bdot\{{\bf X},{\bf X}\}_{\rm gy}\bdot{\bf V} \;=\; -\,\frac{c\bhat_{0}}{qB_{\|}^{*}}\bdot{\bf U}\btimes{\bf V}
\label{eq:UV_id}
\end{equation} 
for any two vector fields $({\bf U},{\bf V})$.

Cyclic permutations of the Interaction sub-bracket \eqref{eq:I} include the second-order terms
 \begin{eqnarray}
&&{\bf F}\bdot\left\{{\bf X},\frac{}{}{\bf K}\bdot\{{\bf X},g\}_{\rm gy}\right\}_{\rm gy} \;-\; {\bf K}\bdot\left\{{\bf X},\frac{}{}{\bf F}\bdot\{{\bf X},g\}_{\rm gy}\right\}_{\rm gy} \nonumber \\
 &&=\; F_{i}\,K_{j} \left( \left\{ \{g, X^{j}\}_{\rm gy},\frac{}{} X^{i}\right\}_{\rm gy} + \left\{ \{X^{i}, g\}_{\rm gy},\frac{}{} X^{j}\right\}_{\rm gy} \right) \nonumber \\
  &&+\; \left({\bf F}\bdot\{{\bf X},{\bf K}\}_{\rm gy} \;-\frac{}{} {\bf K}\bdot\{{\bf X},{\bf F}\}_{\rm gy}\right) \bdot\{{\bf X},g\}_{\rm gy},
  \label{eq:I2}
\end{eqnarray}
where the last two terms can be expressed as
\begin{equation}
\left(\frac{c\bhat_{0}}{qB_{\|}^{*}}\btimes{\bf K}\bdot\nabla{\bf F} \;-\; \frac{c\bhat_{0}}{qB_{\|}^{*}}\btimes{\bf F}\bdot\nabla{\bf K}\right)\bdot\{{\bf X}, g\}_{\rm gy}.
\label{eq:I2_id}
\end{equation} 
Hence, by combining Eqs.~\eqref{eq:V2_id} and \eqref{eq:I2_id}, we obtain
\begin{widetext}
\begin{eqnarray}
\{{\bf X}, g\}_{\rm gy}\vb{\cdot}\left[ \left(\frac{c\bhat_{0}}{qB_{\|}^{*}}\vb{\times}{\bf F}\right)\vb{\times}\nabla\vb{\times}{\bf K} - \left(\frac{c\bhat_{0}}{qB_{\|}^{*}}\vb{\times}{\bf K}\right)\vb{\times}\nabla\vb{\times}{\bf F}\right] &=& \frac{c\bhat_{0}}{qB_{\|}^{*}}\vb{\cdot}\left[ \nabla\vb{\times}{\bf K}\, ({\bf F}\vb{\cdot}\{{\bf X},g\}_{\rm gy}) - ({\bf F}\vb{\cdot}\nabla\vb{\times}{\bf K})\,\frac{{\bf B}^{*}}{B_{\|}^{*}}\,\pd{g}{p_{\|}}\right] \nonumber \\
 &&-\; \frac{c\bhat_{0}}{qB_{\|}^{*}}\vb{\cdot}\left[ \nabla\vb{\times}{\bf F}\, ({\bf K}\vb{\cdot}\{{\bf X},g\}_{\rm gy}) - ({\bf K}\vb{\cdot}\nabla\vb{\times}{\bf F})\,\frac{{\bf B}^{*}}{B_{\|}^{*}}\,\pd{g}{p_{\|}}\right],
 \label{eq:VI2_id}
 \end{eqnarray} 
where we used the Poisson-bracket identity $(c\bhat_{0}/qB_{\|}^{*})\bdot\{{\bf X}, g\}_{\rm gy} = (c/q B_{\|}^{*})\,\partial g/\partial p_{\|}$.

Cyclic permutations of the Maxwell sub-bracket \eqref{eq:M} include the second-order terms
\begin{equation}
-\,\frac{c\bhat_{0}}{qB_{\|}^{*}}\vb{\cdot}\left[ \nabla\vb{\times}{\bf K}\, ({\bf F}\vb{\cdot}\{{\bf X},g\}_{\rm gy}) - ({\bf F}\vb{\cdot}\nabla\vb{\times}{\bf K})\,\frac{{\bf B}^{*}}{B_{\|}^{*}}\,\pd{g}{p_{\|}}\right] \,+\, \frac{c\bhat_{0}}{qB_{\|}^{*}}\vb{\cdot}\left[ 
\nabla\vb{\times}{\bf F}\, ({\bf K}\vb{\cdot}\{{\bf X},g\}_{\rm gy}) - ({\bf K}\vb{\cdot}\nabla\vb{\times}{\bf F})\,\frac{{\bf B}^{*}}{B_{\|}^{*}}\,\pd{g}{p_{\|}}\right].
 \label{eq:M2}
 \end{equation}
 \end{widetext}
 which exactly cancel the terms in Eq.~\eqref{eq:VI2_id}, so that the second-order Jacobi term in Eq.~\eqref{eq:Jacobi_0123} is obtained by adding Eqs.~\eqref{eq:V2}, \eqref{eq:I2}, and \eqref{eq:M2}, which yields the second-order term in the Jacobiator \eqref{eq:Jacobi_0123}:
\begin{eqnarray}
{\sf Jac}_{2}[{\cal F},{\cal G},{\cal K}] &=& F_{i}\,K_{j} \left[ \left\{ \{X^{i},g\}_{\rm gy},\frac{}{} X^{j} \right\}_{\rm gy} \frac{}{}\right. \nonumber \\
 &&+\; \left\{ \{X^{j}, X^{i}\}_{\rm gy},\frac{}{} g \right\}_{\rm gy} \nonumber \\
 &&\left.+\frac{}{} \left\{\{g,\; X^{j}\}_{\rm gy},\frac{}{} X^{i}\right\}_{\rm gy}\right] + \leftturn = 0,
 \label{eq:Jacobi_2}
\end{eqnarray} 
which vanishes because of the Jacobi property \eqref{eq:gyPB_Jacobi} of the gyrocenter Poisson bracket \eqref{eq:PB_gy}.

\subsection{Third-order Jacobi property}

At the third order in $4\pi q$, we note that only the Interaction and Maxwell sub-brackets \eqref{eq:I} and \eqref{eq:M} contribute terms in ${\sf Jac}_{3}[{\cal F},{\cal G},{\cal K}]$. First, the third-order terms in cyclic permutations of Eqs.~\eqref{eq:I} and \eqref{eq:M} are
\begin{widetext}
\begin{equation}
{\bf K}\vb{\cdot}\left\{ {\bf X},\frac{}{} {\bf F}\vb{\cdot}\{{\bf X},{\bf X}\}_{\rm gy}\vb{\cdot}{\bf G}\right\}_{\rm gy} + \leftturn \;=\; {\sf Jac}_{3}[{\cal F},{\cal G},{\cal K}] \;+\; \left[ {\bf K}\vb{\cdot}\{{\bf X},{\bf X}\}_{\rm gy}\vb{\cdot}{\bf G} \left(\frac{c\bhat_{0}}{qB_{\|}^{*}}\vb{\cdot}\nabla\vb{\times}{\bf F}\right) \;+\; \leftturn \right], 
\label{eq:I3}
\end{equation}
and
\begin{equation}
{\bf G}\vb{\cdot}\{{\bf X},{\bf X}\}_{\rm gy}\vb{\cdot}{\bf K} \left(\frac{c\bhat_{0}}{qB_{\|}^{*}}\vb{\cdot}\nabla\vb{\times}{\bf F}\right) \;+\; {\bf K}\vb{\cdot}\{{\bf X},{\bf X}\}_{\rm gy}\vb{\cdot}{\bf F} \left(\frac{c\bhat_{0}}{qB_{\|}^{*}}\vb{\cdot}\nabla\vb{\times}{\bf G}\right)
\;+\; {\bf F}\vb{\cdot}\{{\bf X},{\bf X}\}_{\rm gy}\vb{\cdot}{\bf G} \left(\frac{c\bhat_{0}}{qB_{\|}^{*}}\vb{\cdot}\nabla\vb{\times}{\bf K}\right),
\end{equation}
\end{widetext}
which when combined, using the antisymmetry property of Eq.~\eqref{eq:UV_id}, introduces simple cancellations and yields the result
\[ {\bf K}\vb{\cdot}\left\{ {\bf X},\frac{}{} {\bf F}\vb{\cdot}\{{\bf X},{\bf X}\}_{\rm gy}\vb{\cdot}{\bf G}\right\}_{\rm gy} + \leftturn \;=\; {\sf Jac}_{3}[{\cal F},{\cal G},{\cal K}], \]
where the third-order term in the Jacobiator \eqref{eq:Jacobi_0123}:
\begin{eqnarray}
{\sf Jac}_{3}[{\cal F},{\cal G},{\cal K}] &=& F_{i}\,G_{j}\,K_{\ell} \left[ \left\{ \{X^{i},X^{j}\}_{\rm gy},\frac{}{} X^{\ell} \right\}_{\rm gy} \frac{}{}\right. \nonumber \\
 &&+\; \left\{ \{X^{j}, X^{\ell}\}_{\rm gy},\frac{}{} X^{i} \right\}_{\rm gy} \nonumber \\
 &&\left.+\frac{}{} \left\{\{X^{\ell},\; X^{i}\}_{\rm gy},\frac{}{} X^{j}\right\}_{\rm gy}\right] = 0,
 \label{eq:Jacobi_3}
\end{eqnarray} 
vanishes because of the Jacobi property \eqref{eq:gyPB_Jacobi} of the gyrocenter Poisson bracket \eqref{eq:PB_gy}.

\section{\label{sec:summary}Summary}

The explicit Hamiltonian structure of the gauge-free gyrokinetic Vlasov-Maxwell equations was constructed directly from Eqs.~\eqref{eq:V_bracket}-\eqref{eq:E_bracket} in terms of a gyrokinetic Hamiltonian functional \eqref{eq:Ham_gy} and the gyrocenter Poisson bracket \eqref{eq:PB_gy}, which resulted in the gyrokinetic Vlasov-Maxwell bracket \eqref{eq:gyVM_bracket}. The gauge-free gyrokinetic equations were presented here in their drift-kinetic limit, which simplified the expressions for the gyrocenter polarization and magnetization \eqref{eq:P_gy}-\eqref{eq:M_gy}. Future work will consider extensions of our gyrokinetic Hamiltonian formulation to include higher-order FLR effects.

As simple applications of our gauge-free gyrokinetic Hamiltonian formulation, we demonstrated in Sec.~\ref{sec:bracket} that the gyrokinetic entropy function \eqref{eq:S_gy} is a Casimir functional for the gyrokinetic Vlasov-Maxwell bracket \eqref{eq:gyVM_bracket}, and that the gyrokinetic toroidal angular momentum conservation law can be expressed in Hamiltonian form \eqref{eq:P_phi_Ham}. 

In Sec.~\ref{sec:Jacobi}, we presented an explicit proof that the gyrokinetic Vlasov-Maxwell bracket \eqref{eq:gyVM_bracket} satisfies the Jacobi property \eqref{eq:Jacobi}. While this may seem to be an academic exercise, our proof follows similar proofs for 
several Vlasov-Maxwell models presented in the Appendix of Ref.~\cite{Morrison_2013}, for example. In analogy to the recent more extensive proof \cite{Brizard_proof_2021} of the Jacobi property of the guiding-center Vlasov-Maxwell bracket, the proof of the Jacobi property \eqref{eq:Jacobi} relies on identities derived from the gyrocenter Poisson bracket \eqref{eq:PB_gy} and the vanishing gyrokinetic Jacobiator \eqref{eq:Jacobi} is inherited from the Jacobi property \eqref{eq:gyPB_Jacobi} of the gyrocenter Poisson bracket.

Future work will consider an alternate gauge-free gyrokinetic Vlasov-Maxwell model \cite{Brizard_2021} in which the gyrocenter polarization drift $\epsilon\,\vb{\Pi}_{1}\bdot d{\bf X}/dt$ is added to the symplectic part of the gyrocenter Lagrangian \eqref{eq:Lag_gy}, where $\vb{\Pi}_{1} \equiv [{\bf E}_{1} + (p_{\|}\bhat_{0}/mc)\btimes{\bf B}_{1}]\btimes q\bhat_{0}/\Omega_{0}$. This extension will explicitly introduce electric-field terms in the gyrocenter Poisson bracket \eqref{eq:PB_gy} and introduce new terms in the Poisson variation \eqref{eq:delta_Poisson}.

\acknowledgments

The Author acknowledges useful discussions with J.W.~Burby and Eero Hirvijoki, and thanks the Referees for their careful review. This work was supported by the National Science Foundation grant No.~PHY-1805164.

\vspace*{0.1in}

\begin{center}
{\bf Data Availability Statement}
\end{center}

Data sharing is not applicable to this article as no new data were created or analyzed in this study.

\appendix

\section{Gyrocenter angular momentum equation of motion}

In this Appendix, we derive the equation of motion for the gyrocenter angular momentum \eqref{eq:P_phi} in an axisymmetric background magnetic field. 

We begin with the covariant azimuthal (toroidal) component of the Euler-Lagrange equation \eqref{eq:EL_X}:
\begin{eqnarray}
 &&\epsilon q\left({\bf E}_{1} + \frac{1}{c}\frac{d{\bf X}}{dt}\btimes{\bf B}_{1}\right)\bdot\pd{\bf X}{\varphi} - \pd{K_{\rm gy}}{\varphi} \nonumber \\
  &&=\; \frac{dp_{\|}}{dt}\,\left(\bhat_{0}\bdot\pd{\bf X}{\varphi}\right) + \frac{q}{c}{\bf B}_{0}^{*}\btimes\frac{d{\bf X}}{dt}\bdot\pd{\bf X}{\varphi},
  \label{eq:A1}
 \end{eqnarray}
 where the toroidal derivative of the gyrocenter kinetic energy $\partial K_{\rm gy}/\partial\varphi = (\partial{\bf X}/\partial\varphi)\bdot\nabla K_{\rm gy}$ is given in Eq.~\eqref{eq:K_phi}. 
 
 Next, we explicitly evaluate the gyrocenter time derivative of $P_{\varphi}$:
 \begin{eqnarray}
 \frac{dP_{\varphi}}{dt} &=& \frac{dp_{\|}}{dt}\,\left(\bhat_{0}\bdot\pd{\bf X}{\varphi}\right)  \nonumber \\
  &&+\; \frac{q}{c}\frac{d{\bf X}}{dt}\bdot\left[\nabla{\bf A}_{0}^{*}\bdot\pd{\bf X}{\varphi} + \nabla\left(\pd{\bf X}{\varphi}\right)\bdot{\bf A}_{0}^{*}\right] \nonumber \\
  &=& \frac{dp_{\|}}{dt}\,\left(\bhat_{0}\bdot\pd{\bf X}{\varphi}\right) + \frac{q}{c}{\bf B}_{0}^{*}\bdot\left(\frac{d{\bf X}}{dt}\btimes\pd{\bf X}{\varphi}\right) \nonumber \\
   &&+\; \wh{\sf z}\btimes{\bf A}_{0}^{*}\bdot\frac{q}{c}\frac{d{\bf X}}{dt} \;+\; \wh{\sf z}\bdot\left(\frac{d{\bf X}}{dt}\btimes\frac{q}{c}\,{\bf A}_{0}^{*}\right) \nonumber \\
    &=& \frac{dp_{\|}}{dt}\,\left(\bhat_{0}\bdot\pd{\bf X}{\varphi}\right) + \frac{q}{c}{\bf B}_{0}^{*}\btimes\frac{d{\bf X}}{dt}\bdot\pd{\bf X}{\varphi},
 \end{eqnarray}
where we used the gyrocenter invariance of the magnetic moment $\mu$ and we used $\partial{\bf A}_{0}^{*}/\partial\varphi \equiv \wh{\sf z}\btimes{\bf A}_{0}^{*}$ under the assumption of axisymmetry of the background magnetic field, while we used the identity ${\bf W}\bdot\nabla(\partial{\bf X}/\partial\varphi)\bdot{\bf V} \equiv \wh{\sf z}\bdot({\bf W}\btimes{\bf V})$, which holds for arbitrary vectors $({\bf V},{\bf W})$. Hence, Eq.~\eqref{eq:A1} yields the gyrocenter angular momentum equation of motion
 \begin{equation}
 \frac{dP_{\varphi}}{dt} \;=\; \epsilon q\left({\bf E}_{1} + \frac{1}{c}\frac{d{\bf X}}{dt}\btimes{\bf B}_{1}\right)\bdot\pd{\bf X}{\varphi} - \pd{K_{\rm gy}}{\varphi},
 \label{eq:P_dot}
 \end{equation}

 We note that in the electrostatic limit, where ${\bf E}_{1} = -\,\nabla\Phi_{1}$ and ${\bf B}_{1} = 0$, Eq.~\eqref{eq:P_dot} yields the standard equation $dP_{\varphi}/dt = -\,\partial H_{\rm gy}/\partial\varphi$, where $H_{\rm gy} = \epsilon q\,\Phi_{1} + K_{\rm gy}$.

\end{document}